\documentclass[a4paper,onecolumn,11pt,english]{article}
\usepackage[a4paper,top=2.54cm,bottom=2.54cm,left=2.54cm,right=2.54cm]{geometry}
\usepackage{fancyhdr}
\usepackage{tgbonum}
\usepackage{mathrsfs,amsmath} 
\usepackage{float}
\usepackage{lipsum}
\usepackage{subfig}
\usepackage{subfloat}
\usepackage{wrapfig}
\usepackage{graphicx}
\usepackage{pslatex}
\usepackage{amsmath}
\usepackage{amsfonts}
\usepackage{setspace}
\usepackage{epstopdf}
\usepackage{wrapfig}
\usepackage[numbers,sort&compress]{natbib} 
\usepackage[table,dvipsnames]{xcolor}
\usepackage{csquotes}
\usepackage{hyperref}
\usepackage{graphicx}

\makeatletter
\usepackage{babel}
\makeatother
\fontfamily{phv}
\graphicspath{{../article_patches/figures/}}

\let\OLDthebibliography\thebibliography
\renewcommand\thebibliography[1]{
  \OLDthebibliography{#1}
  \setlength{\parskip}{0pt}
  \setlength{\itemsep}{0pt plus 0.3ex}
}

\begin{document}

\title{\textbf{Demolishing prejudices to get to the foundations: a criterion of demarcation for fundamentality} }
\author{\\Flavio Del Santo$^{1,2,3}$ and Chiara Cardelli$^{1,2}$ \\ \\ \begin{small}\textit{$^{1}$Faculty of Physics, University of Vienna, Boltzmanngasse 5, Vienna A-1090, Austria} \end{small}\\  \begin{small} \textit{
 $^{2}$Vienna Doctorate School of Physics (VDS)}\end{small}\\	\begin{small} \textit{
 $^{3}$Basic Research Community for Physics (BRCP)}\end{small}}

\date{}

    \maketitle
    \begin{abstract}
    \begin{small}
   In this paper, we reject commonly accepted views on fundamentality in science, either based on bottom-up construction or top-down reduction to isolate the alleged \emph{fundamental entities}. We do not introduce any new scientific methodology, but rather describe the current scientific methodology and show how it entails an inherent search for foundations of science. This is achieved by phrasing (minimal sets of) metaphysical assumptions into falsifiable statements and define as fundamental those that survive empirical tests. The ones that are falsified are rejected, and the corresponding philosophical concept is demolished as a prejudice. Furthermore, we show the application of this criterion in concrete examples of the search for fundamentality in quantum physics and biophysics.
      \end{small}\vspace{0,5cm}
      \end{abstract}

%\tableofcontents
\setcounter{page}{1}
%\clearpage
\section{Introduction}

Scientific communities seem to agree, to some extent, on the fact that certain theories are more fundamental than others, along with the physical entities that the theories entail (such as elementary particles, strings, etc.). But what do scientists mean by fundamental? This paper aims at clarifying this question in the face of a by now common scientific practice. We propose a criterion of demarcation for fundamentality based on (i) the formulation of metaphysical assumptions in terms of falsifiable statements, (ii) the empirical implementation of crucial experiments to test these statements, and (iii) the rejection of such assumptions in the case they are falsified.  Fundamental are those statements, together with physical entities they define, which are falsifiable, have been tested but not been falsified, and are therefore regarded as fundamental limitations of nature. Moreover, the fundamental character of statements and entities is strengthened when the same limitations are found through different theoretical frameworks, such as the limit on the speed of propagation imposed by special relativity and independently by no-signaling in quantum physics. This criterion adopts Popper's falsificationism to turn metaphysical concepts into falsifiable statements but goes beyond this methodology to define what is fundamental in science. Such a criterion, as well as Popper's falsificationism, are not here regarded as normative, but rather descriptive of the methodology adopted by a broad community of scientists.

The paper is structured as follows: in Section~\ref{1} we explore the reasons to go beyond reductionism in the search for fundamentality. In  Section~\ref{2} we show that many physicists are employing a falsificationist method, or at least are convinced to do so, which inevitably shapes their research programs. In  Section~\ref{3},  we define our criterion for fundamentality and show its application to notable examples of modern foundations of quantum mechanics --wherein the no-go theorems (such as Bell's inequalities) play a pivotal role--  and in biophysics, in the search for fundamental properties of polymers.

\section{Reductionism and foundations}
\label{1}
Tackling the question of ``what is fundamental?'' seems to boil down, in one way or another, to the long-standing problem of reductionism. This is customarily intended to mean ``that if an entity $x$ reduces to an entity $y$ then $y$ is in a sense \emph{prior to $x$}, is \emph{more basic than $x$}, is such that $x$ \emph{fully depends upon} it or is \emph{constituted by it}'' \cite{sep-scientific-reduction}. Accordingly, the process of reduction is sometimes thought to be equivalent to the action of \emph{digging into} the foundations of science.\footnote{A more realistic-inclined thinker would rather prefer foundations of ``nature''. We will discuss the problem of realism in what follows, but showing that it is an unnecessary concept for our discussion on fundamentality.}
Despite this generally accepted view, we show that the reductionist approach to foundations, which seems \emph{prima facie} legitimate given its historical success, can be overcome by more general approaches to search for foundations.

Reductionism commonly ``entails realism about the reduced phenomena'' \cite{sep-scientific-reduction}, because every effect at the observational level is reduced to objectively existing microscopic entities. This is the case of a particular form of reductionism known as microphysicalism.
 In this view, the entities are the ``building blocks'' of nature, and their interactions fully account for all the possible natural phenomena. \footnote{Philosophy of science discerns between at least three forms of reductionism: theoretical, methodological and ontological. The distinction, although essential, turns out to be quite technical, and it goes beyond the scope of the present essay to expound it. In what follows, we will consider reductionism mainly in the classical sense of microphysicalism (i.e. a mixture of theoretical and ontological reduction). In his book Hüttemann~\cite{hutterman} explores the meaning and limitations of microphysicalism advocating a view in which ``neither the parts nor the whole have 'ontological priority'''.}.This is, however, a view that requires the higher-order philosophical pre-assumption of realism. \footnote{Although the problem of realism is indeed one of the most important in philosophy of science and surely deserves a great deal of attention, it is desirable that this problem finds its solution within the domain of science (this will be discussed in the Subsection \ref{subsec:quantum}).}

Reductionism is justified merely on historical arguments, that is, looking at ``specific alleged cases of successful reductions'' \cite{sep-scientific-reduction}. However, there are many cases where reductionism has exhausted its heuristic power, and it is only the unwarranted approach of some physicists who regard physics as the foremost among sciences, claiming that every biological or mental process can be eventually reduced to mere physical interactions. Feynman, for instance, would maintain that ``everything is made of atoms. [...] There is nothing that living things do that cannot be understood from the point of view that they are made of atoms acting according to the laws of physics'', \cite{feynman1}, Sect. 1.9. On the contrary, we believe, with David Bohm, that ``the notion that everything is, in principle, reducible to physics [is] an unproved assumption, which is capable of limiting our thinking in such a way that we are blinded to the possibility of whole new classes of fact and law'', \cite{bohm}.\footnote{The concept of the impossibility in principle to reduce all event to the underlying physics is not new in philosophy of science, and it usually goes under the name of \emph{multiple-realizability}. To this extent, it is worth mentioning the influential work of the Nobel Laureate in Physics, P. W. Anderson, \cite{anderson}.} Moreover, the reductionist program seems to be failing even within physics alone, not having so far been capable of unifying the \emph{fundamental} forces nor its most successful theories (quantum and relativistic physics). It has been proposed that even a satisfactory theory of gravity requires a more holistic  (i.e. non-reductionist) approach \cite{pollo} and it could have an emergent origin \cite{niels}. Furthermore, it is the belief of many contemporary scientists (especially from the promising field of complex systems studies) that \emph{emergent} behaviours are inherent features of nature \cite{kim}, not to mention the problem of consciousness. So, the authoritative voice of the Stanford Encyclopedia of Philosophy concludes that ``the hope that the actual progress of science can be successfully described in terms of reduction has vanished'' \cite{sep-scientific-reduction}. 

Another tempting path to approach the question of ``what is fundamental'', is the use of conventionalist arguments. ``The source of conventionalist philosophy would seem to be wonder at the austerely beautiful \emph{simplicity of the world} as revealed in the laws of physics'', \cite{popper}, p. 80. The idea, however, that our descriptions being \emph{simple}, \emph{elegant}, or \emph{economical} or the like, constitute a guarantee of ``fundamentality'' is a mere utopia. Conventionalism, despite being totally self-consistent, fails when it comes to acquiring empirical knowledge. In a sense, for the conventionalist, the theory comes first, and observed anomalies are ``reabsorbed'' into \emph{ad hoc} ancillary hypotheses. It thus appears quite unsatisfactory to address foundations of natural science from the perspective of something that has hardly any empirical content.

In conclusion,  although we acknowledge that an approach that involves the intuitive decomposition of systems in \textit{basic building-blocks} of nature can still be a useful heuristic tool, it seems too restrictive to be used  to define a universal criterion of fundamentality. Nor it looks promising to rely on purely conventional (e.g. aesthetic) factors, though they can be fruitful in non-empirical sciences. Indeed, while reduction-based foundations clash with the \emph{ontological} problem (the assumption of realism), conventional-based foundations clash with the \emph{epistemological} problem (the empirical content of theories). What we are left with is to go back to the very definition of science, to its method, and try to understand what science can and cannot do.

\section{Scientists adhere to falsificationism}
\label{2}
As it is generally known, Karl R. Popper \cite{popper} showed the untenability of a well-established criterion of demarcation between science and non-science based on inductive \emph{verification}.\footnote{In particular, Popper's criticisms were leveled against the logical positivism of the Vienna Circle. He indeed came back to Hume's problem of induction; Hume maintained that there is, in fact, no logically consistent way to generalize a finite (though arbitrarily large) number of single empirical confirmations to a universal statement (as a scientific law is intended to be). Popper embraced this position, but he proposed a new solution to the problem of induction and demarcation (see further).}
Popper proposed instead that theories are conjectures that can only be  (deductively) \emph{falsified}. Popper's criterion of demarcation between science and non-science requires that scientific statements (laws, consistent collections of laws, theories) ``can be singled out, by means of empirical tests, in a negative sense: \emph{it must be possible for an empirical scientific system to be refuted by experience}. [...] Not for nothing do we call laws of nature `laws': the more they prohibit the more they say'', \cite{popper}, p. 40-41.

We ought to stress, however, that one of the major critiques to Popper's falsificationism is that it demarcates scientific statements from non-scientific ones on a purely logical basis, i.e. in principle independently of the practical feasibility. In fact, for Popper, a statement is scientific if and only if it can be formulated in a way that the set of its possible falsifiers (in the form of singular existential statements) is not empty. On this regard, \v{C}. Brukner and M. Zukowski, who significantly contributed to FQM in recent years, slightly revised Popper's idea (among many others). While maintaining a falsificationist criterion of demarcation, they attribute to falsifiability a momentary value:
\begin{displayquote}
$[$Non-scientific] propositions could be defined as those which are not observationally or experimentally
falsifiable at the given moment of the development of human knowledge.\footnote{We have preferred here to change the original quotation from ``philosophical'' to ``non-scientific'' inasmuch we do not limit the extent of philosophical to be considered antithetical to scientific statements.}
\end{displayquote}

Falsificationism seems to influence the working methodology of scientists directly. In fact, within the domain of scientific (i.e. falsifiable) statements, we show here that scientists aim at devising crucial experiments to rule out those that will be falsified. In the following, we show that this is employed by many scientists as a scientific methodology. In fact, as the eminent historian of science Helge Kragh recently pointed out, ``Karl Popper's philosophy of science [...] is easily the view of science with the biggest impact on practising scientists'' \cite{kragh}. For instance, the Nobel laureate for medicine, Peter Medawar, acknowledged to Popper's \emph{falsificationism} a genuine descriptive value, stating that ``it gives a pretty fair picture of what actually goes on in real-life laboratories'' \cite{medawar}. Or the preeminent cosmologist Hermann Bondi declared that ``there is no more to science than its method, and there is no more to its method than Popper has said'' \cite{jammer}. Besides these appraisals, it is a matter of fact that ``many scientists subscribe to some version of simplified Popperianism'' \cite{kragh}, and this happens especially to physicists, and specifically to those who are concerned with fundamental issues. 

In this section, we will support this claim with several quotations from different prominent physicists who do not share a common philosophical standpoint, and show that they do actually think of their scientific praxis as based on a form of deductive hypothesis-testing-falsification process.
We then show that this methodological choice has indeed profound consequences on the development of theories and that it has been extremely efficient in the modern results of foundations of different branches of physics.

Here, we are not concerned with the justification of falsificationism as the \emph{right} methodology to aspire to; we avoid any normative judgment.\footnote{It is well known that Popper's methodology has today hardly any supporter among philosophers of science, who have severely criticized it as a too strict and naive description of scientific development. Moreover, falsificationism has for Popper a normative value, i.e. it is seen as the most rational, and therefore the best, possible methodology. Among Popper's foremost critics and commentators, we ought to mention Imre Lakatos, who developed a weaker and more sophisticated form of falsificationism, which encompasses part of Thomas Kuhn's critiques.} We just assume as a working hypothesis - build upon a number of instances - that this is what scientists do, or at least what they are convinced to do: this is enough to lead them to pursue certain (theoretical) directions.  Methodological rules are a matter of convention. They are indeed intuitively assumed by scientists in their everyday practical endeavour, but they are indispensable meta-scientific (i.e. logically preceding scientific knowledge) assumptions: ``they might be described as the rules of the game of empirical science. They differ from rules of pure logic rather as do the rules of chess''. However, a different choice of the set of rules would necessarily lead to a different development of scientific knowledge (meant as the collection of the provisionally acknowledged theories). The methodology that one (tacitly) assumes entails the type of development of scientific theories, insofar as it ``imposes severe limitations to the kind of questions that can be answered'' \cite{feynman}. Our fundamental theories look as they do also because they are derived under a certain underlying methodology. This is not surprising: if chess had a different set of rules (e.g. pawns could also move backwards), at a given time after the beginning of a game, this would probably look very different from any game played with  standard rules of chess. 

Coming back to the physicists who, more or less aware of it, loosely adhere to falsificationism, we deem it interesting to explicitly quote some of them, belonging to different fields. It is worth mentioning a work by the Austrian physicist Herbert Pietschmann with the significant title: ``The Rules of Scientific Discovery Demonstrated from Examples of the Physics of Elementary Particles''. Elementary particle physics was particularly promoted in the Post-war period to revive scientific (especially European) research. It is well known that this field developed in a very pragmatic and productivist way (see e.g.~\cite{io1}). Nevertheless, the author shows that falsificationist methodological rules
\begin{displayquote}
are applied by the working physicist. Thus these rules are shown to be actual tools rather
than abstract norms in the development of physics. [...]
Predictions by theories and their tests by experiments form the basis of the work of scientists. It is common knowledge among scientists that new predictions are not proven by experiments, but are ruled out if they are wrong. \cite{pietschmann}
\end{displayquote}
While Pietschmann has a vast knowledge of the philosophy of science, one of the most brilliant physicist of all times, the Nobel laureate Richard P. Feynman, was rather an ignoramus in philosophy. Feynman belonged to a generation of hyper-pragmatic American scientists, whose conduct went down in history with the expression ``shut up and calculate!'' (see e.g. \cite{mermin}). However, in the course of some public lectures he gave in the 1960s, Feynman's audience was granted the rare opportunity to hear the great physicist addressing the problem of the scientific method. It turns out that he also adheres to falsificationism:
\begin{displayquote}

[scientific] method is based on the principle that observation is the judge of whether something is so or not. [...] Observation is the ultimate and final judge of the truth of an idea. But ``prove" used in this way really means ``test," [...] the idea really should be translated as, ``The exception tests the rule." Or, put another way, ``The exception proves that the rule is wrong." That is the principle of science. \cite{feynman}
\end{displayquote}

Coming to some leading contemporary figures in the field of foundations of quantum mechanics (FQM), David Deutsch has been a staunched Popperian since his student years. Deutsch maintains that 
\begin{displayquote}
we do not read [scientific theories] in nature, nor does nature write them into us. They are
guesses – bold conjectures. [...] However, that was not properly understood until [...] the work of the philosopher Karl Popper \cite{deutsch}.
\end{displayquote}

To conclude, we agree with Bohm when he states that ``scientists generally apply the scientific method, more or less intuitively'' \cite{bohm}. But we also maintain that since scientists are both the proposers and the referees of new theories, the form into which these theories are shaped is largely entailed by the method they (more or less consciously) apply. Methodology turns therefore into an active factor for the development of science. As we will show in the next section, the falsificationist methodology, vastly adopted in modern physics, has opened new horizons for the foundations of physics.\footnote{We must stress that falsificationism is not always considered as the driving scientific method, especially in other scientific disciplines than physics. For instance, Roald Hoffmann, chemist and philosopher of chemistry, claims that falsification is much less relevant to chemistry than to physics, especially when chemists synthesize molecules: ``Synthesis is a creative activity, and while every synthesis implicitly and trivially tries to falsify some deep-seated fundamental law, the science and art of synthesis as a whole does not explicitly and non-trivially try to falsify any particular theory. That does not mean that falsification is absent or untrue, it just means that it is rather irrelevant in this field''~\cite{{hoffman}}}.

\section{A criterion for fundamentality}
\label{3}
\begin{figure*}[h!]
\centering
\includegraphics[width=14cm]{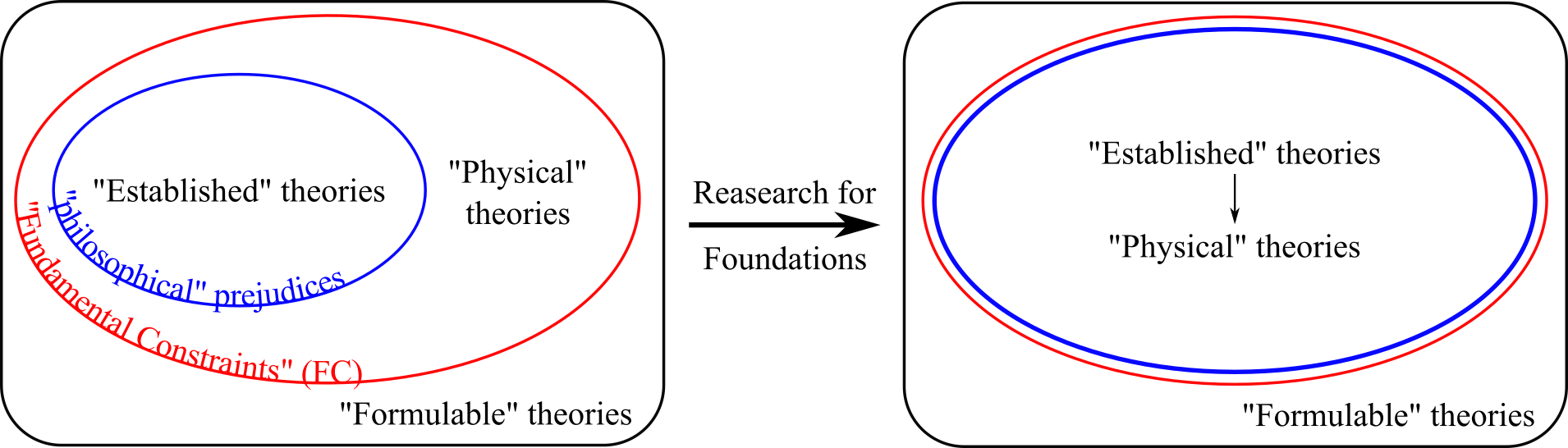}
\caption{\footnotesize{A sketch of fundamental research here proposed as a series of successive experimental violations of the ``philosophical prejudices'' assumed by our ``established theories'', towards the actual fundamental constraints (see main text). ``Physical theories'' here means ``physically significant'', i.e. they carry an empirical content. ``Formulable'' theories are in general all the theories one can think of, and they are characterized solely by the formalism.}}
\label{bop}
\end{figure*}

Provided with a working methodology, we can now propose a criterion to define what is fundamental.  At a naive stage of observation, our intuitive experience leads to the conviction that  such as determinism, simultaneity, realism, were \textit{a priori} assumptions of a scientific investigation. What it turns out, however, is that there is in principle no reason to pre-assume anything like that. The process of reaching the foundations consists of: (i) the process of turning those metaphysical concepts into scientific (i.e., falsifiable) statements, thus transferring them from the domain of philosophy to the one of science (e.g., from locality and realism one deduces Bell's inequalities); (ii) a \textit{pars destruens} that aims at rejecting the metaphysical assumptions, of which the corresponding scientific statements have been tested and empirically falsified. Those rejected metaphysical assumptions were to be considered ``philosophical prejudices''.\footnote{Notice that the expression ``philosophical prejudices'' is borrowed from Feyerabend \cite{feyerabend} (see main text). However, we would like to stress that would be perhaps more correct to call these ``metaphysical assumptions''.} Modern physics, with the revolutionary theories of quantum mechanics and relativity, has washed away some of them, and recent developments are ruling out more and more of these prejudices. Feyerabend's words sound thus remarkable when he states that it
\begin{displayquote}
becomes clear that the discoveries of quantum theory look so surprising only because we were caught in the philosophical thesis of determinism [...]. What we often refer to as a “crisis in physics” is also a greater, and long overdue, crisis of certain philosophical prejudices.” \cite{feyerabend}
\end{displayquote}
We are so able to put to the test concepts that were classically not only considered part of the domain of philosophy (metaphysics) but even necessary \textit{a priori} assumptions for science. Reaching the foundations of physics then means to test each of these concepts and remove the constraints built upon a prejudicial basis, pushing the frontier of scientific domain up to the ``actual'' insurmountable constraints which demarcate the possible from the impossible (yet within the realm of science on an empirical basis, see Fig. \ref{bop}). These actually \textit{fundamental  constraints} (FC) are thus trans-disciplinary and should be considered in every natural science. Yet, contrarily to the physicalist program, the search for FC does not elevate one particular science to a  leading, \emph{more fundamental} position. Moreover, this criterion does not entail any pre-assumption of realism, but rather it allows to test, and possibly empirically falsify, certain forms of realism (see further).  

We would like to clearly point out that, although this criterion aims at removing metaphysical \textit{a priori} assumptions (here called ``philosophical prejudices'') by means of empirical refutations, it is not our aim --and it would not be possible-- to remove philosophy from scientific theories. First of all, as explained above, methodological rules are necessarily non-scientific, and yet they are indispensable for science. Secondly, and this could admittedly  be a problematic issue for our proposed criterion, there is always room for more than one interpretation of empirical evidence. In fact, the process of turning ``philosophical prejudices'' into a falsifiable statement usually introduces further independent metaphysical assumptions. The best that can be done within a theory is to minimize the number of independent ``philosophical prejudices'' from which the falsifiable statement can be deduced. If the falsification occurs, however, only the conjunction of such assumptions is refuted and there is, in general, no way to discriminate whether any of them is individually untenable. Nevertheless, other philosophical consideration can bridge this gap, giving compelling arguments for refuting one subset or the other of the set of ``philosophical prejudices'' (see section \ref{subsec:quantum} for a concrete example on Bell's inequality).

To overcome one possible further criticism to our view, we would like to clarify that not all the falsifiable statements of a scientific theory that have not yet been falsified are to be considered fundamental. Fundamental is a particular set of falsifiable statements that have been deduced from metaphysical assumptions, have been tested, and resisted empirical falsification. Hence, the limitations imposed by no-go results are (provisionally) to  be considered fundamental.
%%%%%%%%%%%%%%%%%%%%%%%%%%%%%%%%%%%%%%
On the same note, a separate discussion deserve the \emph{postulates} of scientific theories, to which intuitively one would like to attribute a fundamental status. A postulate of a scientific theory can be defined as an element of a minimal set of assumptions from which it is possible to derive the whole theory. Note that a postulate may not be directly falsifiable, but at least its conjunction with additional assumptions have to lead to falsifiable statements. Admittedly, one could object that if a postulate did not rest upon any ``philosophical prejudices'', then it could not be fundamental according to our proposed criterion. However, we maintain that it is, in fact, always possible to identify one or more philosophical prejudices that underlie any statement and, in particular, any postulate. This has been discussed thoroughly in the literature, for instance, Northrop states:
\begin{displayquote}
Any theory of  physics makes more physical and philosophical assumptions than the facts alone give or imply. [...] These assumptions, moreover, are philosophical in character.
They may be ontological, i.e., referring to the subject matter of
scientific knowledge which is independent of its relation to the
perceiver; or they may be epistemological, i.e., referring to the
relation of the scientist as experimenter and knower to the
subject matter which he knows. \cite{northrop}
\end{displayquote}
Therefore, the postulates of the scientific theories that we consider fundamental (e.g., quantum physics or special relativity) are fundamental according to our criterion.\footnote{Consider the example of one of the postulates of quantum mechanics which states: ``To any physically measurable quantity of physics is associated a Hermitian operator''. For how technical and abstract this postulate may seem, it ensures that observable quantities take value in the domain of real numbers. Thus, if by any experimental procedure it were possible to ``observe'' a complex number as the outcome of a physical experiment, this would mean a violation of the philosophical prejudice of ``reality of the physical observables''. Being this based on a clear philosophical assumption and since the falsifiable statements that this imply are hitherto not falsified, this postulate clearly is fundamental according to our proposed criterion.}

We can push our criterion of fundamentality even further and try to reach the most fundamental constraints, those that are not provisional because violating them would jeopardize the applicability of the scientific method. A clear example of this is the bound imposed by the finite speed of light. But why is it so special? Why should be the impossibility of instantaneously signaling considered a more fundamental (i.e. insurmountable) limitation of physics, than the bound imposed by, say, determinism? Because the knowledge of the most fundamental bounds is also limited by our methodology. Since falsificationism requires some ``cause-effect'' relations to test theories meaningfully, then instantaneous signaling would break this possibility and any meaning of the current methodology along with it.\footnote{Recent lines of research show that quantum mechanics may not present a definite causal order \cite{costa}. However, these new possibilities do not undermine testable cause-effect relations, and thus do not clash with falsifiability.}

This criterion allows also to attempt a ranking of relative fundamentality between theories, which is alternative to the physicalist approach in terms of building blocks decomposition. We say that a theory $T_1$ is thus more fundamental than a theory $T_2$ when $T_1$ includes $T_2$, but it comes out of the boundaries of $T_2$ (e.g. $T_1=\ $quantum theory and $T_2=\ $deterministic contextual theories; or local realistic theories). In this sense, quantum mechanics is more fundamental than classical mechanics. The reason for this, however, is not that we believe, as it is usually assumed, that the equations quantum mechanics can predict everything that classical mechanics can also predict (in fact, we cannot solve the Schr{\"o}dinger equation for atoms with more than a few electrons). On the contrary, we call quantum physics more fundamental than classical physics because it violates certain no-go theorems (thus demolishing some ``philosophical prejudices''), and yet it lies within the bounds of the empirically tested no-go theorems.

To sum up, we maintain that one of the aims of science is to approach the actual foundations (red edge in Fig. \ref{bop}) through a discrete process of successive falsifications of the alleged \emph{a priori} metaphysical assumptions, which have to be formulated in terms of scientific statements. If these are falsified, then former can be dismissed as ``philosophical prejudices''.

\subsection{Foundations of quantum mechanics}\label{subsec:quantum}

A large part of the research on modern foundations of physics has developed along with the directions that we have thus far described. It is the case of what are usually referred to as ``no-go theorems''.\footnote{Historically, no-go theorems are associated with the conditions of compatibility with quantum formalism. Here, however, we consider them in a more general context. They are regarded as decidable statements that discriminate between any two classes of theories, possibly beyond QM (see further).} They require a \emph{falsifiable statement} (e.g. in the form of an inequality) that is deductively inferred (i.e. formally derived) from a minimal set of assumptions, which are chosen to include the ``philosophical prejudices'' (in the sense expounded above) that one wants to test. A no-go theorem, therefore, allows formulating one or more of this ``philosophical prejudices'' in terms of a statement that can undergo an experimental test: what till then was believed to be a philosophical assumption, suddenly enters the domain of science. If this statement is experimentally falsified, its falsity is logically transferred to the conjunction of the assumptions (\emph{modus tollens}) that thus becomes untenable. The no-go theorem is the statement of this untenability.

But there is more to the epistemological power of ``no-go theorems'': they can sometimes be formulated in a way that they do not include any particular scientific theory in their assumptions (\emph{device-independent} formulation). In this case, the no-go theorem assumes the form of a collection of measurements and relations between measurements (\emph{operational} formulation), yet it holds directly independently of any specific experimental apparatus, its settings, or the chosen degrees of freedom to be measured.   
In practice, if a particular theory assumes one of the ``philosophical prejudices'' that have been falsified by a certain no-go theorem, then this theory needs to be revised (if not completely rejected) in the light of this evidence. Furthermore, the falsification of a no-go theorem rules out the related ``philosophical prejudice'' for every future scientifically significant theory.

We shall review some of the by now classical no-go theorems in quantum theory, in the spirit of the present paper. It is generally known that quantum mechanics (QM) provides only probabilistic predictions, that is, given a certain experiment with measurement choice $x$ and a possible outcome $a$, quantum theory allows to compute the probability  $p(a|x)$ of finding that outcome. Many eminent physicists (Einstein, Sch\"{o}dinger, de Broglie, Bohm, Vigier, etc.) made great efforts to restore \emph{determinism} and \emph{realism}.\footnote{Realism is a vague concept, but it can be summarised as the idea that measurements reveal pre-existing properties of the system. However, in the context of no-go theorems, this should be formally defined from time to time.} A way to achieve this is to assume the existence of underlying \emph{hidden variables} (HV), $\lambda$, not experimentally accessible (either in principle or provisionally), that if considered would restore determinism, i.e. $p(a|x,\lambda)=0$ or $1$.\footnote{This is more of a historical motivation, than an actual program. Hidden variables, in fact, do not necessarily have to be deterministic. What it really matters for HV theories is that there exist a \emph{complete} state encompassing the ``real state of affairs'', i.e. the values of all the physical properties independently of any measurements. Strict determinism was given up also by Bohm himself.} In a celebrated work \cite{bohm2}, Bohm proposed a fully developed model of QM in terms of HV. However, the HV program started encountering some limitations. To start with, S. Kochen and E. Specker \cite{ks} assumed (1) a deterministic HV description of quantum mechanics and (2) that these HV are independent of the choice of the disposition of the measurement apparatus (context),\footnote{In principle, and classical physics assumes this, one can use whatever practical method to measure a certain physical quantity and  always find the same value (within experimental uncertainty).} and showed that this leads to an inconsistency.\footnote{It is noteworthy that the original proof by Kochen and Specker involved an at least three-dimensional Hilbert space and  117 directions of projection. This can be seen as rather \emph{unsimple} and \emph{inelegant}, as a practical example against conventionalist arguments.} Thus if HV exist, they must depend on the context. John Bell, however, noticed that this is not so surprising since
\begin{displayquote}
there is no a priori reason to believe that the results [...] should be the same. The result of observation may reasonably depend not only on the state of the system (including hidden variables) but also on the complete disposition of apparatus. (\cite{bell66}, p. 9)
\end{displayquote}
We must stress that this theorem rules out the conjunction of the assumptions only \emph{logically}. It is only with an experimental violation, recently achieved \cite{ks exp}, of its falsifiable formulation that contextuality is ruled out.

But it was with a seminal paper by Bell \cite{bell64}, that one of the most momentous no-go theorems was put forward.\footnote{To explain this result we follow a more recent \emph{a posterior} reconstruction of Bell's theorem (e.g. \cite{brunner}, \cite{brukner1}).} 
Consider two distant (even space-like separated) measurement stations $A$ and $B$. Each of them receives a physical object (\textit{information carrier}) and they are interested in measuring the correlations between the two information carriers that have interacted in the past. At station $A$ ($B$) a measurement is performed with settings labeled by $x$ ($y$), and the outcome by $a$ ($b$). Since the stations are very far away and the local measurement settings are freely chosen, common sense (or a ``philosophical prejudice'') would suggest that the joint probability of finding $a$ and $b$ given $x$ and $y$ is independent (i.e. factorizable). Nevertheless, in principle (i.e. without ``prejudices''), the local measurement settings could somehow statistically influence outcomes of distant experiments, such that $p(a, b|x,y) \neq   p(a|x) p(b|y)$. It is important to notice that "the existence of such correlations is nothing mysterious. [...] These correlations may simply reveal some dependence relation between the two systems which was established when they interacted in the past" \cite{brunner}. This 'common memory' might be taken along by some \textit{hidden variables} $\lambda$ that, if considered, would restore the independence of probabilities. The joint probability then becomes:\footnote{$\lambda$ can, in general, be governed by a probability distribution and be a continuous variable over a domain $\Lambda$, as considered below. The final probability $p(a,b|x,y)$ should eventually not explicitly depend on $\lambda$, which should be averaged out.}
 \begin{equation}
 p(a,b|x,y) = \int_{\Lambda}d \lambda \ q(\lambda) p(a|x,\lambda)p(b|y,\lambda).
\label{lr}
 \end{equation}
This condition is referred to as \textit{local realism} (LR).\footnote{The name comes from the fact that decomposition \eqref{lr} was derived under the mere assumption of having some \emph{real} quantities $\lambda$ that factorize the joint probability distribution into \emph{local} operations only. Notice, however, that LR is here a compound condition, given by the mathematical expression \eqref{lr}, and cannot be formally separated into two distinct conditions as some authors try to do (e.g. \cite{brunner}) to justify the ``non-local'' nature of QM (see further).}

Let us consider dichotomic measurement settings and outputs (i.e. $x$, $y \in \{0,1\}$ and $a$, $b \in \{-1, +1\}$) and define the correlations as the averages of the products of outcomes given the choices of settings, i.e. $\langle a_xb_y \rangle = \sum_{a, b } ab   \   p(a,b|x,y)$, it is easy to prove that the condition \eqref{lr} of LR leads to the following expression in terms of correlations: 
 \begin{equation}
S_{(LR)} = \langle a_0b_0 \rangle + \langle a_0b_1 \rangle+\langle a_1b_0 \rangle-\langle a_1b_1 \rangle\leq 2.
\label{chsh}
 \end{equation}
This is an extraordinary result, known as \emph{Bell's inequality} \cite{bell64}.\footnote{In fact, this is the easiest non-trivial Bell's inequality, usually referred to as Clause-Horne-Shimony-Holt (CHSH) inequality. Bell's inequalities can be generalized to an arbitrary number of settings and outcomes, and to many parties.} Indeed, a condition such as \eqref{lr} gives a mathematical description of the profound metaphysical concepts related to \emph{locality} and \emph{realism}, whereas its derived form \eqref{chsh}, transforms LR into an experimentally falsifiable statement in terms of actually measurable quantities (correlations). Indeed,
\begin{displayquote}
the conjunction of all assumptions of Bell inequalities is not a philosophical statement, as it is testable
both experimentally and logically [...]. Thus, Bell’s theorem removed the question of the possibility of local realistic description from the realm of philosophy. \cite{brukner1}
\end{displayquote}
Since the 1980s, experiments of increasing ambition have \emph{tested} local realism through Bell's inequalities \cite{giustina}, and have empirically \emph{violated} them. Namely, LR has been \emph{falsified} and this removed the possibility of scientific theories based on a local realistic description. It ought to be remarked that Bell's no-go theorem relies on the additional tacit assumption of  'freedom of choice' (or 'free will'): local settings are chosen freely and independently from each other. This is a clear example of the problematic issue previously stated, namely that the conjunction of these metaphysical assumptions (i.e., local realism and freedom of choice) is at stake, but the falsification of the statement deduced from them cannot discriminate which one should be dropped (or if they both are untenable). However, the rejection of the assumption of freedom of choice would mean   that all of our experiments are in fact meaningless, because we would live within a 'super-deterministic' Universe, and the decision of rotating a knob or otherwise would be a mere illusion. Therefore, believing that we are granted the possibility of experimenting is a philosophical argument that makes us reject freedom of choice as a valuable alternative. Hence, in our view, we are entitled to interpret the violation of Bell's inequalities as a refutation of local realism.\footnote{Some scholars maintain that there Bell's theorem requires the further implicit assumption that quantum laws are valid also in those physical situations in which quantum mechanics forbids that a measurement context exists to test their validity \cite{aaa}. If this were the case, a further ambiguity in the ``philosophical prejudice'' to be refuted would again arise. We are thankful to one of the reviewers for providing this reference.}

Quantum mechanics is a compatible theory because its formalism gives a result that is out of the bounds of local realism.
Indeed, quantum mechanics allows the preparation of pairs of information carriers called \emph{entangled}.\footnote{Formally, quantum mechanics postulates that systems are described by vectors $|\psi \rangle$ living in complex Hilbert spaces, $\mathcal{H}$. The joint state of two systems A and B lives in the tensor product of the spaces of the two systems, i.e. $\mathcal{H}_{AB}= \mathcal{H}_{A} \otimes \mathcal{H}_{B}$. A pure state $|\psi \rangle_{AB} \in \mathcal{H}_{AB}$ is then defined to be separable if $|\psi \rangle_{AB}= |\psi \rangle_{A} \otimes |\psi \rangle_{B}$. Otherwise, it is entangled. A maximally entangled state (for two two-level systems) is for instance the \emph{singlet} state: $| \psi^- \rangle =  \frac{1}{\sqrt2}\left( |0\rangle_A \otimes |1\rangle_B - |1\rangle_A \otimes |0\rangle_B \right) )$. The states $|0\rangle$ and $|1\rangle$ are the eigenstates of the standard Pauli $z$-matrix $\sigma_z$ associated to the eigenvalues +1 and -1, respectively}. From elementary calculations (see e.g. \cite{brunner}), it follows that using quantum entanglement, the relation between correlations as defined in \eqref{chsh} reaches a maximum value (Tsirelson's bound) of
 \begin{equation}
S_{(Q)}= 2\sqrt2 > 2 = S_{(LR)}.
\label{quant} 
 \end{equation}
This is the second crucial result of Bell's inequalities: the quantum formalism imposes a new bound that is out of the bounds of local realism. At the moment this new bound imposed by \eqref{quant} has never been experimentally violated, and QM survived experimental falsification.
There is yet another condition that one might want to enforce, namely that the choice of measurement settings cannot directly influence the outcomes at the other stations (not in terms of correlations but actual information transfer). This is called \emph{no-signaling} (NS) condition  and reads 
 \begin{equation}
  \sum_b  p(a,b|x,y) = p (a|x);  \ \ \ \ \  \sum_a  p(a,b|x,y) = p (b|y),
 \end{equation}
The NS constraint is where we set the FC. Indeed, a theory that would violate this condition allows for instantaneous signaling and it thus would mean a failure of the scientific method as we conceive it. It would be in principle \emph{not} falsifiable (besides being incompatible with relativity theory).\footnote{Concerning quantum and no-signaling correlations, it is possible to show that the quantity S of the correlations defined in \eqref{chsh} reaches a maximal logical bound $S_{(NS)}= 4$. S. Popescu and D. Rohrlich \cite{pr} explicitly formulated a set of correlations that reach this logical bound, but that respect NS (i.e. they are physically significant).}
 \begin{figure}[]
\centering
\includegraphics[width=6.5cm]{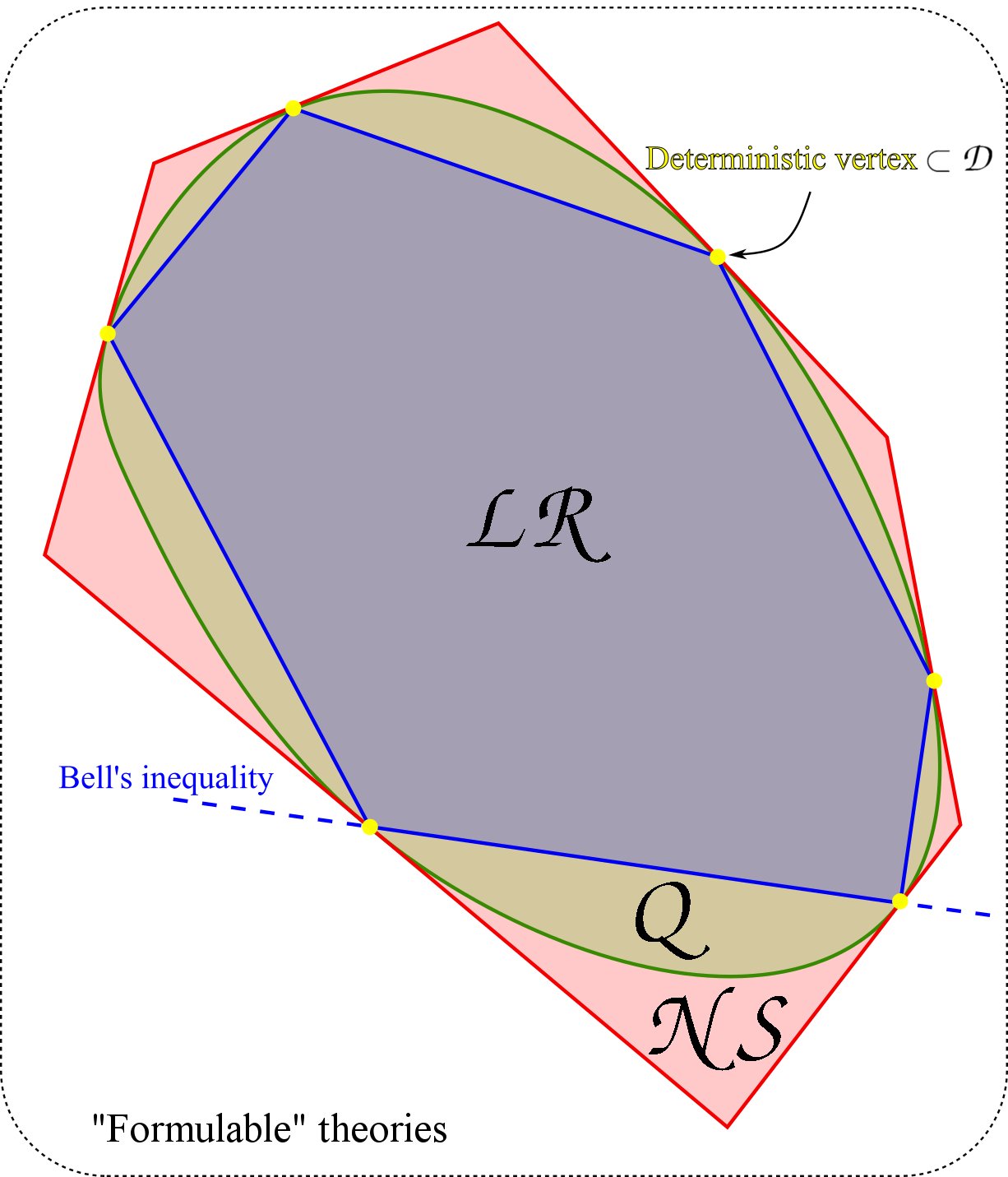}
\caption{\footnotesize{Relations between local realistic ($\mathcal{LR}$), quantum ($\mathcal{Q}$) and no-signaling ($\mathcal{NS}$) theories in a probability space. Although here the geometry is also explicit (see footnote \ref{aaa}), the strict inclusion $ \mathcal{LR} \subset \mathcal{Q} \subset \mathcal{NS} \subset$ ``formulable'' theories, represents the series of bounds abstractly sketched in Fig. \ref{bop}.}}
\label{gatto1}
\end{figure}

It is possible to show that LR correlations are a proper subset of quantum correlations and that both are strictly included in the NS set of theories (see Fig. \ref{gatto1}).\footnote{\label{aaa}In general, for every number of values that $x$, $y$, $a$, $b$ can take, it is possible to prove that in the space of all the possible probabilities $p(ab|xy)$, the LR condition \eqref{lr} forms a \emph{polytope} whose vertexes are the deterministic correlations ($\mathcal{D}$), and whose facets (the edges in the 2-d representation of Fig. \ref{gatto1}) are the Bell's inequalities (Minkowski's theorem assures that a polytope can be always represented as the intersection of finitely many hyperplanes).  NS correlations form a polytope too, whereas the quantum correlations form a convex, closed and bound set, but that has no facets.} To summarise, local realistic theories have been falsified, and we have a theory, QM, which comes outside its borders. However, it is not the most fundamental theory we think of, since there is potentially room for theories that violate the bounds imposed by QM, and still lies in the domain of ``physically significant'' theories (i.e within the NS bound).

In the literature of modern FQM, there is plenty of other no-go theorems that quantify the discrepancy between ``classical'' and quantum physics, ruling out different ``philosophical prejudices'' than the mentioned contextuality and local realism. For instance, it has been recently proposed \cite{io2} a new no-go theorem for information transmission. %The authors then demonstrate a violation of a bound of exchange of communication, derived under the conjunction of these assumptions.
Consider a scenario in which information should be transmitted between two parties, $A$ and $B$ in a time window $\tau$ that allows a single information carrier to travel only once from one party to the other (``one-way'' communication). At time $t=0$, $A$ and $B$ are given inputs $x$ and $y$ and at $t=\tau$ they reveal outputs $a$ and $b$. The joint probability results in a classical mixture of one-way communications:
\begin{eqnarray}
p(a,b|x,y) =  \gamma p_{A}(a|x)p_{A\prec B}(b|x,y,a)+ \nonumber \\
   + (1-\gamma)p_{B}(b|y)p_{B\prec A}(a|x,y,b),
\end{eqnarray}
where symbol $\prec$ denotes the direction of communication, e.g. $A\prec B$ means that $A$ sends the information carrier to $B$. This distribution leads to a Bell-like inequality that, in the case of $x,y,a,b=0,1$, reads:
 \begin{equation}\label{class dist}
p(a=y,b=x) \leq \frac{1}{2}.
\end{equation}
In Ref.\cite{io2} is shown that an information carrier in quantum superposition between $A$ and $B$ surpasses this bound, and leads to $p(a=y,b=x)= 1$. This bound, logically violated by quantum formalism, has been also experimentally falsified, and results in a violation of ``classical'' one-way communication.

\subsection{``Foundations'' of biophysics}
The process of reaching foundations here proposed can also   be pursued in branches of physics considered more ``complex" (i.e. the opposite of fundamental in the reductionist view), like the physics of biological systems. 

Proteins are a class of polymers involved in most of the natural processes at the basis of life. The 3D structure of each protein, of which the precision is essential for its functioning, is uniquely encoded in a 1D sequence of building blocks (the 20 amino acids) along a polymer chain. This process of encoding is usually referred to as \emph{design}. The huge variability of all existing natural proteins is originated solely by  different sequences of the same set of 20 building blocks. Proteins are very complex systems and the understanding and prediction of the mechanism behind their \textit{folding}, that is the process with which they reach their target 3D structure, is still one of the biggest challenges in science. Until now, no other natural or artificial polymer is known to be designable and to fold with the same precision and variability of proteins.

A more fundamental approach to understand proteins, that can possibly go beyond the observed natural processes, is to ask a different question: are proteins such unique polymers? In other words: is the specific spatial arrangement of the atoms in amino acids the only possible realization to obtain design and folding? \\
According to the mean field theory of protein design~\cite{pande2000}, given an alphabet of building blocks of size $q$, a system is \textit{designable} when 
 \begin{equation}
q>M,
\label{proteins} 
 \end{equation}
 where $M$ is the number of structures that the chain can access, divided the number of monomers along the chain.\footnote{$M$ is $\exp(\omega)$, where $\omega$ is the so-called \textit{configurational entropy} per monomer} For instance, a simple bead and springs polymer will have a certain number of possible structures $\tilde{M}$, defined only by the excluded volume of the beads. If one adds features to the monomers (e.g. directional interactions), this can result in a smaller number of energetically/geometrically accessible structures for the polymer $M<\tilde{M}$. With fewer accessible structures, it will be easier for the sequence to select a specific target structure - and not an ensemble of degenerate structures. Hence, to make a general polymer designable (from Eq. \eqref{proteins}) one can follow two strategies: i) increasing the alphabet size $q$ and, ii) reducing the number of accessible structures per monomer $M$. In the works~\cite{Coluzza2012c,chiara} the authors show, by means of computational polymer models, that one can indeed reduce the number of accessible structures by simply introducing a few directional interactions to a simple bead-spring polymer. This approach is already enough to reduce the number of accessible structures $M$ to obtain designability.\footnote{The designability was obtained for a size of the alphabet $q= 20$ such as in proteins or even less down to $q =3$.} The resulting polymers (\emph{bionic proteins}) can have different numbers and geometries of those directional interactions and, despite not having at all the geometrical arrangement of amino acids, are designable and able to fold precisely into a specific unique target structure, with the same precision of proteins. These bionic proteins are also in principle experimentally realizable on different length scales. Introducing directional interactions is not the only approach to reduce the number of accessible configurations, therefore there can be other examples of polymers that are able to be designable and fold into a target structure. This view teaches us to look beyond the prejudice that the particularism of the amino acids is the only way to achieve design and folding, and search the functioning of proteins into more fundamental principles, that can be applied to a wider range of possible folding polymers.
 \begin{figure}[H]
\centering
\includegraphics[width=6.5cm]{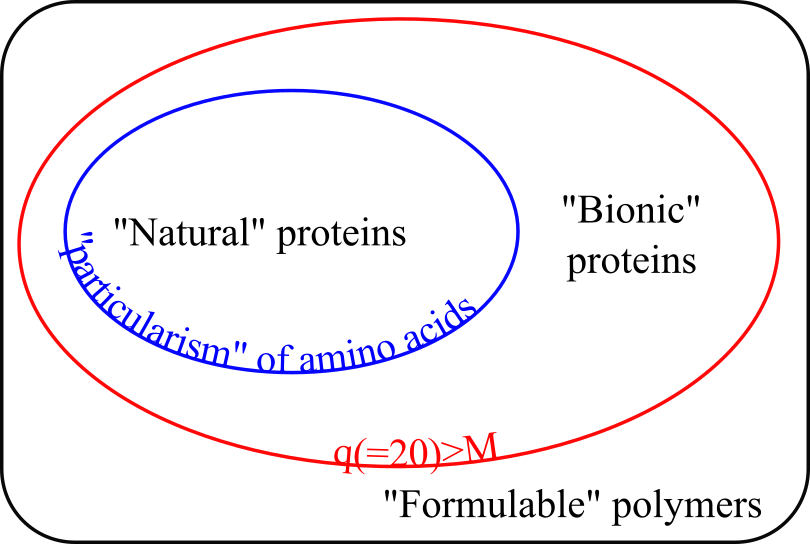}
\caption{\footnotesize{Relations between polymers at a fixed $q=20$: natural proteins, ``bionic'' proteins (i.e. all polymers that are designable to fold into a specific target structure) and all ``formulable'' polymers. The boundary between ``bionic'' proteins and not folding polymers is the inequality \eqref{proteins} (see main text).}}
\label{chiara}
\end{figure}
\section{Conclusions}
We have proposed a criterion for fundamentality as a dynamical process that aims at removing ``philosophical prejudices'' by means of empirical falsification. What is fundamental in our theories are their limits, and these can be discovered performing experiments and interpreting their results. This search tends however to an end, given by the FC (which are the most general, physically significant constraints under a certain methodology).

\textbf{\textit{Acknowledgments}} - The authors thank David Deutsch  for kind comments, and David Miller and \v{C}aslav Brukner for interesting discussions.

\clearpage
\newpage
\begin{small}

\end{small}

\newpage

\end{document}